\documentclass[final,a4paper]{elsarticle}
\usepackage{lineno,enumitem,hyperref}

\usepackage{hyperref,graphicx,amsmath,amsfonts,setspace,enumitem,adjustbox,siunitx}
\usepackage{booktabs,multirow,tikz,color,colortbl,algorithm,algorithmicx,cleveref,csquotes,subcaption}
\usepackage[noend]{algpseudocode}
\usepackage[margin=1in]{geometry}
\graphicspath{{Figures/}{./}}

\usepackage{etoolbox}
\apptocmd{\sloppy}{\hbadness 10000\relax}{}{}

\newcommand{\clrone}{\rowcolor[gray]{.9}[0pt][0pt]}

\renewcommand{\b}{\bfseries}

\renewcommand{\(}{\left(}
\renewcommand{\)}{\right)}

\newcolumntype{L}{l<{\kern\tabcolsep}@{}}
\newcolumntype{C}{c<{\kern\tabcolsep}@{}}
\newcolumntype{R}{r<{\kern\tabcolsep}@{}}
\newcolumntype{M}{r<{\kern-\tabcolsep}}

\begin{document}
\begin{frontmatter}

\title{Exploring the Predictability of Cryptocurrencies via Bayesian Hidden Markov Models}

\author[1]{Constandina Koki\corref{cor1}}
\ead{kokiconst@aueb.gr}
\address[1]{Athens University of Economics and Business, 76 Patission Str. GR-10434 Athens, Greece. }

\author[2]{Stefanos Leonardos}
\ead{stefanos\_leonardos@sutd.edu.sg}
\author[2]{Georgios Piliouras}
\address[2]{Singapore University of Technology and Design, 8 Somapah Rd. 487372 Singapore, Singapore.}
\ead{georgios@sutd.edu.sg}
\cortext[cor1]{Corresponding author}

\begin{abstract}
In this paper, we consider a variety of multi-state Hidden Markov models for predicting and explaining the Bitcoin, Ether and Ripple returns in the presence of state (regime) dynamics. In addition, we examine the effects of several financial, economic and cryptocurrency specific predictors on the cryptocurrency return series. Our results indicate that the Non-Homogeneous Hidden Markov (NHHM) model with four states has the best one-step-ahead forecasting performance among all competing models for all three series. The dominance of the predictive densities over the single regime random walk model relies on the fact that the states capture alternating periods with distinct return characteristics. In particular, the four state NHHM model distinguishes bull, bear and calm regimes for the Bitcoin series, and periods with different profit and risk magnitudes for the Ether and Ripple series. Also, conditionally on the hidden states, it identifies predictors with different linear and non-linear effects on the cryptocurrency returns. These empirical findings provide important insight for portfolio management and policy implementation.

\end{abstract}

\begin{keyword}
Cryptocurrencies \sep Bitcoin \sep Ether \sep Ripple \sep Hidden Markov Models \sep Regime Switching models \sep Bayesian Inference
\JEL C11, C52, E49
\end{keyword}

\end{frontmatter}


\section{Introduction}
\label{sec:introduction}

More than one decade after the successful launch of Bitcoin by the pseudonymous Satoshi Nakamoto \citep{Nak08}, and cryptocurrencies markets are experiencing an enormous growth. At present, there are several thousand cryptocurrencies that constitute a multi-billion market \citep{Hu19}. Due to their increasing popularity and intriguing financial and econometric properties \citep{Dyh18}, they have attracted the interest of investors, regulatory authorities, policymakers, tech-savvy entrepreneurs and academics \citep{Cor19}. Their documented safe haven and hedge properties, most relevant in periods with volatile stock markets and inflationary pressures in fiat currencies, render cryptocurrencies increasingly important in portfolio optimization and risk diversification \cite{Urq19,Bou20,Pla20}. In turn, informed decisions concerning optimal portfolio allocation and asset management require models with good predicting ability \citep{Chen20}. \par
Similar to forecasting studies about conventional financial assets and exchange rates \citep{Mil20,Pan19}, this has prompted a growing literature on the predictability of cryptocurrency returns, Existing results range from the identification of significant explanatory variables (\cite{Aa19,Bl19,Kur19} and \cite{Cor19,Kat19} for comprehensive surveys of earlier models) to price prediction with elaborate machine and deep learning models \citep{Chen20,Che20}. Under the Bayesian framework, main efforts involve continuous state space models \citep{Hot18} univariate and multivariate dynamic linear models, model averaging and time-varying vector autoregression models \citep{Ca19}. These articles show that time-varying models give significantly improved point and density forecasts when compared to various benchmarks such as the random walk model. 
\par 
The improved performance of the state space and time varying models is not a surprise, since these models accommodate various characteristics of the cryptocurrency series, such as time varying volatility and time varying mean returns. Accumulating evidence points to the existence of structural breaks \citep{Me19,Bou19,Kat19,Thi18} return and volatility jumps \citep{She20,Cha18} and regime/state switches \citep{Ar19,Kou19,Ko18} in cryptocurrency returns. However, while regime switching models have been shown to deliver improved forecasting results in exchange rates series and stock market returns (see e.g., \cite{Pan15,Dia15,Yua11} among others), their application in the cryptocurrency context is still limited (cf. \Cref{sub:related}).\par 
Stimulated by the above and aiming to contribute to the growing literature on the predictability of cryptocurrencies, we perform a systematic analysis of various multi-state (regime-switching) Hidden Markov (HM) models on the return series of the three largest (in terms of market capitalization) cryptocurrencies, Bitcoin (BTC), Ether (ETH) and Ripple (XRP). Our goal is to examine the impact of regime switches in predicting the return series and the state-dependent (time-varying) effects of several financial, economic and cryptocurrency specific exogenous predictors. In total, we consider eight discrete state space HM models with exogenous predictors, cf. \Cref{tab:models}. The models include between $2$ and $5$ hidden states,\footnote{To determine the number of states we undertook an extensive specification test. Experiments with more than 5 states (not presented here) exhibit worst performance. Even though adding more states may improve the in-sample fit, the decreased parsimony leads to worse predictions \cite{Gui06}.} and either Homogeneous (HHM) or Non-Homogeneous (NHHM) transition probabilities. We also consider the standard 2-state Markov Switching Random Walk (MS-RW) model without exogenous predictors. We benchmark the aforementioned HM models against three single regime models: the Random Walk (RW) model that is commonly used (as a benchmark) in predicting exchange rates \citep{Pan15,Fro05,Yua11,Che05}, the linear random walk model with all the predictors, often referred to as the \textit{Kitchen Sink} (KS) model, \citep{Ca19}, and the linear Auto-Regressive (AR(5)) model with lagged values up to lag 5. All models are estimated using Bayesian MCMC methods. \par
The predictor set includes exchange rates of various fiat currencies, stock and volatility indices, commodities, and cryptocurrency specific variables. Following the tradition in the literature \citep{Gel14,Ge10,Ber05}, we use the out-of-sample forecasting performance of the aforementioned models to discriminate between the different empirical models. The statistical evaluation of the models is based on the Continuous Rank Probability Score (CRPS) and Mean Squared Forecast Error (MSFE). Finally, to examine if there is an underlying non-linear correlation between the predictors and the return series through the transition probabilities, we add a stochastic search reversible jump step in the NHHM model with the best forecasting performance. We report the posterior probabilities of inclusion in the hidden states transition equations for each predictor. 
\par Our results demonstrate that the 4-states NHHM model has the best forecasting performance for all three series with significant improvements over the single regime models. Even though the optimal model (in terms of predicting accuracy) is common for all three series, we find that the predictors affecting the observed and unobserved processes are significantly different. From the complete set of twelve predictors, the Bitcoin series is affected linearly and non-linearly by five predictors, whereas the Ether and Ripple series are affected by seven predictors. In addition, only the US Treasury Yield and the CBOE stock market volatility index, VIX, have predictive power on all three series. Turning to the in-sample analysis of the 4-state NHHM model, we observe that the returns of each state present distinct characteristics. Unlike conventional exchange rates, see for example \cite{Yua11}, we observe that the hidden states for all cryptocurrencies are not persistent, but they present frequent alternations. In particular, concerning the Bitcoin return series, we find that state 1, the most frequently occurring state, corresponds to a bear regime (i.e., negative returns and high volatility), states 2 and 3 correspond to a bull regime (positive returns and low volatility) but with different kurtosis and state 4 corresponds to a calm regime (returns close to 0 and low volatility). The relation of the BTC returns and volatility within the hidden states is consistent with the asymmetric volatility theory. Regarding the ETH return series, we observe frequent alternations between state 1, the high volatility state, and the low volatility state 2, while states 3 and 4 serve as auxiliary states with low occupancies. Lastly, state 1 of the Ripple series corresponds to periods with extremely high average returns but, as a trade-off, also with high risk. States 2 and 3 are the states with the highest occupancies, while state 4 serves as an auxiliary state.
\par The remainder of the paper is structured as follows. \Cref{sec:DM} provides an overview of the data and methodology. Specifically, \Cref{sec:Data} describes the data along with their transformations and descriptive statistics and \Cref{sec:Methodology} presents the Hidden Markov models and forecasting evaluation criteria. \Cref{sec:Empirical} presents the empirical findings of the forecasting exercise, i.e., the out-of-sample results and the in-sample analysis of the model with the best performance. Finally, \Cref{sec:Conclusion} summarizes and discusses our results. 

\subsection{Other Related Literature}\label{sub:related}
Our model relates to two strands of literature. From a methodological perspective, our model draws from the econometric literature of HM models. Since the seminal work of \cite{Ha89}, HM models have been fruitfully applied in diverse areas such as communications engineering and bioinformatics \cite{Cap06}. In finance, they have been extensively used in predicting and explaining exchange rates \cite{Pan15,Lee06,Fro05,Bol00}, stock market returns \cite{Dia15,Ang09,Gui06}, business cycles \cite{Tia19,Cha06}, realized volatility \cite{Kok20,Liu18}, the behavior of commodities \cite{Per16} and in portfolio allocation \cite{Pla19}. The reason for their increased popularity is that they present various attractive features. In particular, the time-varying parameters which are driven by the state variable of the presumed underlying Markov process, lead to models that can accommodate both non-linearities and mean reversions \cite{Wu14,Gui08}. In addition, HM models can act as filtering processes that account for outliers and abrupt changes in financial market behavior \cite{DiPe16,Ang12} and flexibly approximate general classes of density functions \cite{Tim00}.
\par 
In the cryptocurrency context, HM models have been applied by \cite{Hot18,Ca19} as a state space model, by \cite{Ph17} in the understanding of price bubbles and by \cite{Ko18,Kou19} in examining the relation of BTC with conventional financial assets. \cite{Bou19,Cap19,Ar19} use the HM setting to study volatility of cryptocurrencies with GARCH models. 
The motivation of these studies lies in the observed features of the cryptocurrencies return and volatlity series. In particular, cryptocurrencies series are non-stationary and present non-normalities , heteroskedasticity, volatility clustering, heavy tails and excess kurtosis \cite{Kat17,Kat19,Haf18}. \cite{Cha18} and \cite{Thi18} (among others) document the existence of abrupt price changes and outliers, while \cite{Cor18} show that BTC returns are characterized by an asymmetric mean reverting property. With the aforementioned attractive features of HM models and the characteristics of the cryptocurrency series, it is only natural to ask: do HM models offer improved predictive performance of cryptocurrency returns?

\section{Data and Methodology}\label{sec:DM}
\subsection{The Data}\label{sec:Data}
\begin{figure}[!htb]
\centering
\includegraphics[width=\linewidth,clip =true, trim =4.3cm 1cm 4cm 1.1cm]{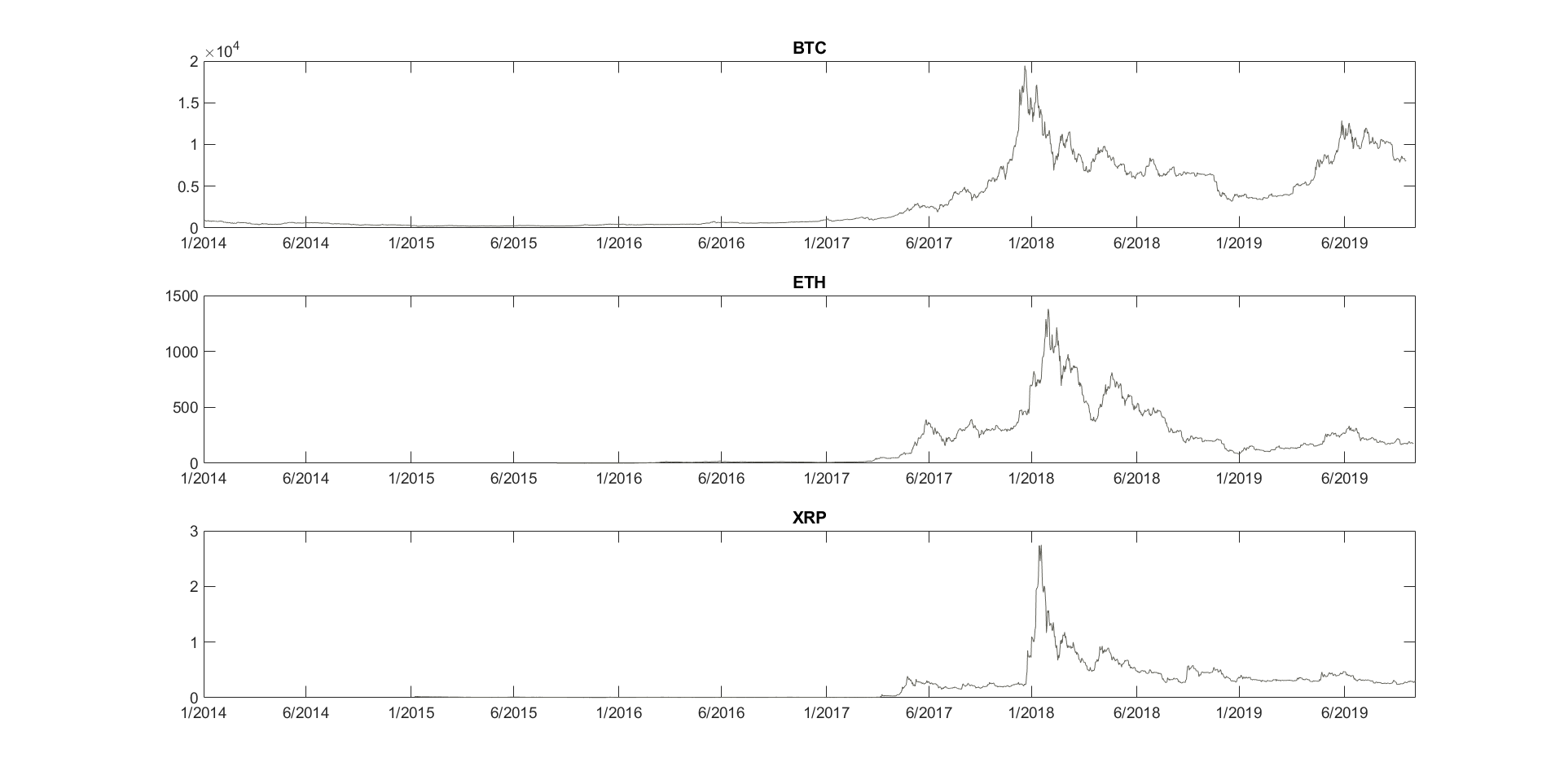}
\caption{Daily price series plots for the three cryptocurrencies considered in this study: Bitcoin (upper plot), Ether (middle plot) and Ripple (lower plot). We study the Bitcoin time series between 1/2014 and 11/2019, the Ether time series between 9/2015 and 11/2019 and the Ripple time series between 1/2015 and 11/2019.}
\label{fig:prices}
\end{figure}

We use the percentage logarithmic end-of-the-day returns, defined as $y_t=100\times \(\log\(p_t\)-\log\(p_{t-1}\)\)$, where $p_t$ denotes the prices of BTC, ETH and XRP. For each coin, we exclude an initial adjustment market period. In particular, we study the BTC time series for the period ranging from 1/2014 until 11/2019, the ETH series for the period ranging from 9/2015 until 11/2019 and the XRP data series from 1/2015 until 11/2019. \Cref{fig:prices} displays the series under study.
The covariate set comprises (i) normalized fiat currencies: Euro to US Dollar (EUR/USD), Great Britain Pound to US Dollar (GBP/USD), Chinese Yuan to US Dollar (CNY/USD) and Japanese Yen to US Dollar (JPY/USD), (ii) commodities: Gold and crude Oil normalized future prices, (iii) stock indices: Standard and Poor's 500 logarithmic returns (SP500), CBOE volatility logarithmic index (VIX), (iv) interest rates: US 10-year Treasury Yield (TY) logarithmic returns and (v) cryptocurrency specific variables: the blockchain block size (SIZE) as percentage of difference between two consecutive days and the percentage of difference between two consecutive days of Hash Rate (HR)\footnote{The Hash rate is used in BTC and ETH series only since XRP is not mineable.}. We report some illustrative descriptive statistics of our covariate set in \Cref{tab:Summary Statistics}.\footnote{The cryptocurrency price series and blockchain size were downloaded from \href{https://www.coinmetrics.io/}{coinmetrics.io}, the exchange rates and commodities prices were downloaded from \href{https://www.investing.com/}{investing.com}, the S\&P 500 index, VIX and Treasury Yield from \href{https://finance.yahoo.com/}{Yahoo Finance} and lastly the Hash Rate from \href{https://www.quandl.com/}{quandl.com} for BTC and from \href{https://etherscan.io}{etherscan.io} for ETH.}  Finally, we also include the lagged 1 autoregressive term of the studied series as a predictive variables. \footnote{The importance of including autoregressive terms is highlighted in \cite{Tim00} who shows that this generates cross-product terms that enhance the set of third- and fourth-order moments and the patterns in serial correlation and volatility dynamics of these models.}
In our experimental study, we added up to 5 lagged autoregressive terms as predictive variables. We did not observed any improvement in the performance and hence we omit the results of these experiments.

\begin{table}[!htb]
\centering
\setlength{\tabcolsep}{10pt}
\arrayrulecolor{gray}
\resizebox{\textwidth}{!}{%
\begin{tabular}{@{}LRRRRRRRRRM}
\hline \clrone\multicolumn{11}{@{}C@{}}{} \\[-0.3cm]
\clrone
\b Variables& \b Transformed &\b Mean  & \b Std. & \b Min.&\b q05&\b p50&\b q95&\b Max.&\b Kurt & \multicolumn{1}{@{}r@{}}{\b Skew}\\
\hline \\[-0.3cm]
\b Bitcoin&\textit{\% log returns}&0.11&3.95&-24.37&-6.47&0.17&6.13&22.47&7.84&-0.27\\
\b Ether&\textit{\% log returns}&0.33&6.54&-31.67&-9.76&-0.01&11.61&30.06&6.64&0.07\\
\b Ripple&\textit{\% log returns}&0.14&7.02&-63.65&-8.46&-0.31&10.25&100.85&38.99&2.55\\
\midrule
\b EUR/USD &\textit{normalized}&0&1&-1.42&-1.20&-2.24&2.05&6.30&6.16&1.50\\
\b GBP/USD&\textit{normalized}&0&1&-5.57&-1.65&0.03&1.64&7.23&8.02&0.15\\
\b CNY/USD&\textit{normalized}&0&1&-16.60&-1.52&0.02&1.57&6.27&41.26&-2.00\\
\b JPY/USD&\textit{normalized}&0&1&-9.70&-1.52&0.04&1.44&6.42&15.13&-0.59\\
\b Gold&\textit{normalized}&0&1&-5.32&-1.60&0&1.61&9.32&9.80&0.29\\
\b Oil&\textit{normalized}&0&1&-7.93&-1.22&0&1.29&10.79&23.92&1.04\\
\b SP500&\textit{log returns}&0&0.01&-0.05&-0.01&0&0.01&0.05&9.47&-0.56\\
\b VIX&\textit{log prices}&2.68&0.25&2.21&2.31&2.63&3.15&3.71&3.37&0.70\\
\b TY&\textit{log returns}&0&0.01&-0.10&-0.03&0&0.03&0.11&7.18&0.13\\
\midrule
\b BTC Hash&$\%$ \textit{of change}&-1.12&14.77&-138.86&-25.42&-0.72&20.81&66.80&10.58&-0.88 \\
\b ETH Hash&$\%$ \textit{of change}&0.42&4.02&-25.50&-4.21&0.37&4.82&99.90&248.92&9.50\\
\b BTC size&$\%$ \textit{of change}&-0.82&13.61&-81.26&-23.04&-0.27&20.84&45.08&5.10&-0.46\\
\b ETH size&$\%$ \textit{of change}&-0.52&14.63&-359.96&-16.87&-0.15&15.81&55.28&243.51&-10.24\\
\b XRP size&$\%$ \textit{of change}&-0.83&13.76&-90.94&-24.63&-0.06&20.43&51.02&6.61&-0.73\\
\bottomrule
\end{tabular}
}
\caption{Summary Statistics of the percentage logarithmic return cryptocurrency series and transformed predictors. The first two columns show the abbreviated name and the transformation of each variable. Columns three to four display the mean and standard deviation of each variable and columns five to nine display the minimum values, the 5\%, 50\%, 95\% quantiles and the maximum values respectively. The last two columns report the kurtosis and skewness coefficients.}
\label{tab:Summary Statistics}
\end{table}

\subsection{The Econometric Framework: Hidden Markov Models}\label{sec:Methodology}
In this study, we focus on Hidden Markov (HM) models. In the HM setting, the probability distribution of the studied series, $Y_t$, depends on the state of an unobserved (hidden) discrete Markov process, $Z_t$. Let $\(Y_t, X_t\)$ denote the pair of the random process of the assumed cryptocurrency return series, $Y_t$, with realization $y_t$, and the set of explanatory variables (predictors), $X_t$, with realization $x_t=\(x_{1t},\dots,x_{kt}\)$, where $k$ denotes the number of predictors. The hidden process, $Z_t$, follows a first order finite Markov process with $m<\infty$ states and transition probabilities $P\(Z_{t+1}=j\mid Z_{t}=i\)=p_{ij}$, $i,j=1,\dots,m$. If $m=1$, then the model is the standard linear regression model. Given the realized state, $Z_t=z_t$, the observed process is modeled as $y_t=g(z_t)+\epsilon_t$, with $g$ a predetermined function and $\epsilon_t$ the associated errors terms. \par
\begin{figure}[!htb]
\centering\vspace{-1cm}
\includegraphics[scale=0.5]{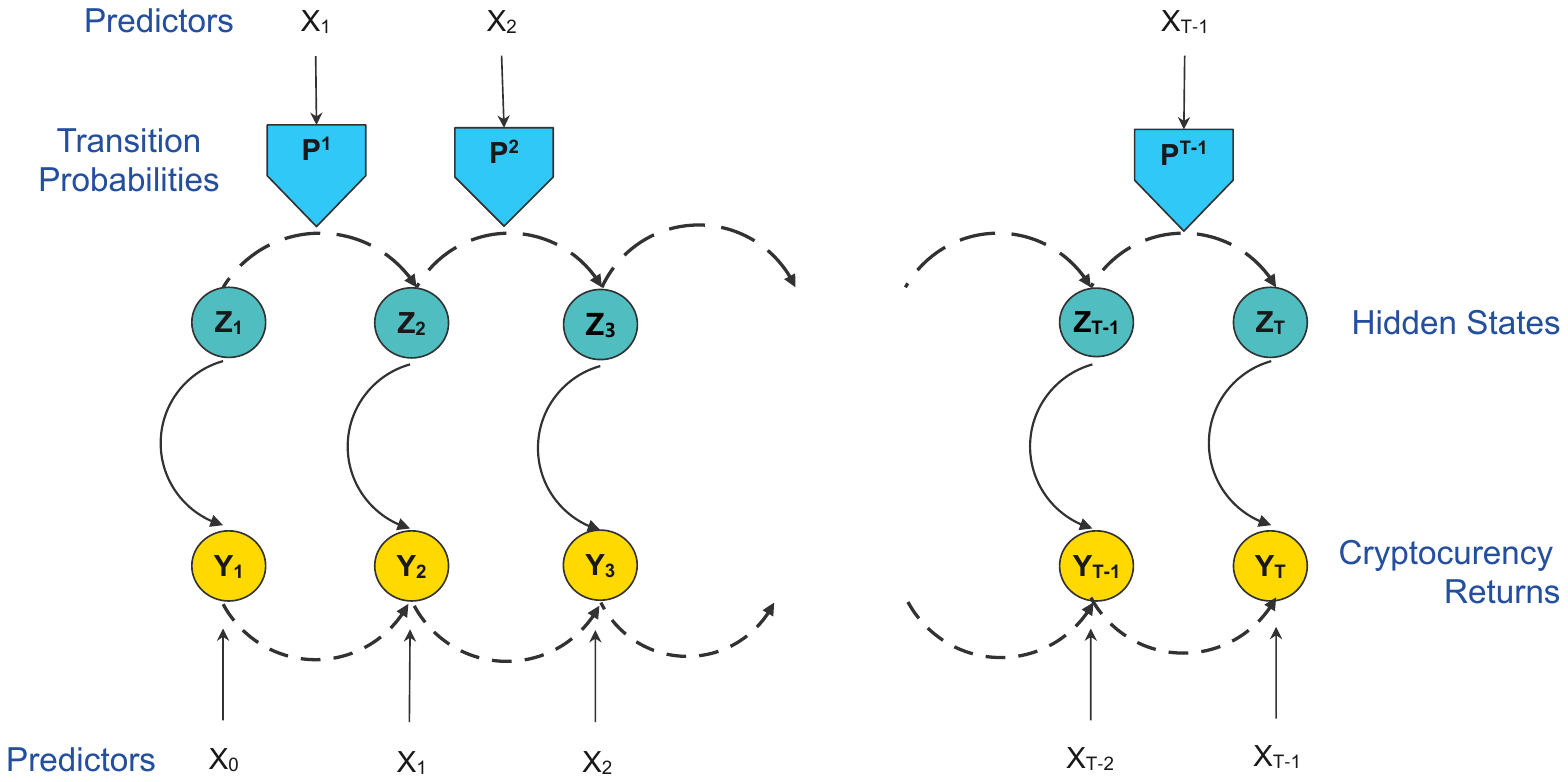}
\vspace{-2cm}
\caption{Graphical representation of the Non-Homogeneous Hidden Markov model.}
\label{fig:Diagram}
\end{figure}

 
We consider the Normal HM models, i.e., we choose a linear function $g$ and normal regression errors. In particular, the observed process can be written in the form, 
$$y_t=B_{z_t}X_t+\epsilon_{Z_t},$$
where $B_{z_t}=\(b_{0z_t},b_{1z_t},\dots,b_{kz_t}\)$ are the regression coefficients when the latent process at time $t$ is at state $z_t=s$, $s=2,\dots,m$, and $\epsilon_{z_t}$ are the normally distributed random shocks, $\epsilon_{z_t}\sim \mathcal{N}\(0,\sigma^2_{z_t}\)$. The hidden process is determined by the transition probability matrix
$$P^{\(t\)} = 
\begin{bmatrix}
p^{\(t\)}_{11} & p^{\(t\)}_{12} & \cdots & p^{\(t\)}_{1m} \\
p^{\(t\)}_{21} & p^{\(t\)}_{2,2} & \cdots & p^{\(t\)}_{2m} \\
\vdots  & \vdots  & \ddots & \vdots  \\
p^{\(t\)}_{m1} & p^{\(t\)}_{m2} & \cdots & p^{\(t\)}_{mm} 
\end{bmatrix},
$$
where $p^{\(t\)}_{ss'}=P\(Z_{t+1}=s'\mid Z_t=s\)$ is the probability that at time $t+1$ the hidden state is $s'$ given that at time $t$ the hidden state was $s$. If the transition probabilities are time-constant, then the resulting model is a \textit{Homogeneous Hidden Markov} (HHM) model. If we relax this hypothesis, then the resulting model is the more flexible \textit{Non-Homogeneous Hidden Markov} (NHHM) model. A graphical representation of the NHHM is shown in  \Cref{fig:Diagram}.
To model the transition probabilities, we use a multinomial link with predictors $X_t$ and multinomial regression coefficients $\beta _{ij}=\(\beta _{0,ij},\beta _{1,ij},\dots ,\beta_{k,ij}\)^{\prime }$
$$p^{(t)}_{ij}=\dfrac{\exp\(x_{t}\beta_{ij}\)}{\displaystyle\sum^{m}_{l=1}\exp\(x_t\beta_{il}\)}, \; i,j=1,\dots,m.$$ 
Based on the P\'{o}lya-Gamma data augmentation scheme of \cite{Po13}, this model uses a latent variable scheme to make inference on the multinomial regression coefficients which leads to more accurate and robust estimations \citep{Kok20}.\par
To make inference on the models' parameters we use a Bayesian Markov Chain Monte Carlo (MCMC) algorithm, which comprises the following steps: (a) A Forward-Backward (FB) algorithm for simulating the hidden states \citep{Sc02}, 
(b) a Gibbs sampling step for estimating the linear regression coefficients using conditional conjugate analysis, (c) a Gibbs sampling step with a P\'{o}lya-Gamma data augmentation scheme for estimating the multinomial regression coefficients and (d) simulation of $L$ one-step-look-ahead forecasts. To study the effects of the predictors on the non-homogeneous transition probabilities, we interpolate between the fourth (c) and fifth (d) steps a stochastic variable search (via reversible jump) step.
\par In addition, we include in our analysis the 2-state Markov-Switching Random Walk (MS-RW) model with drift \cite{Eng94,Rzh12} and Normal errors, as a simpler and parsimonious regime switching model. This allows the drift term $\mu$ and the variance $\sigma^2$ to take two distinct values, i.e., 
$$y_t\sim \mathcal{N}\(\mu_{z_t},\sigma^2_{z_t}\) , \text{} z_t=1,2,$$ where the variable $Z_t$ is governed by the constant transition probabilities $P(Z_t =1\mid Z_{t-1}=1)=p_{11}$ and $P\(Z_t =2\mid Z_{t-1}=2\)=p_{22}$. 

\subsection{Bayesian Forecasting and Performance Evaluation}
We assess the performance of the studied models based on their predicting ability. Reflecting the logical positivism of the Bayesian approach stating that \emph{a model is as good as its predictions} 
\cite{Ge10,Gui11}, the predictive accuracy is valued not only for its own sake but rather for comparing different models within the Bayesian framework.  
\par In this study, we obtain a sample of the $L$ one-step-look ahead predictions, $\hat{Y}_l, l=1,\dots,L$, from the empirical posterior predictive distribution, utilizing the MCMC output. Specifically, at the $i$-th MCMC iteration and for $l=1\dots L$ we construct an one-step-look-ahead forecast $\hat{y}_{l,i}$ as follows. Given the model parameter estimates, we simulate the hidden state $Z_l$ form the discrete distribution based on the transition probability matrix $P^{\(l-1\)}_i$. Then, conditional on the hidden state, we draw the prediction $\hat{y}_{l,i}$ from the $\mathcal{N}\(X_{l-1}B_{Z_{l}},\sigma^2_{Z_l}\)$. Thus, by the end of the MCMC iterations, we get a sample of the $L$ empirical predictive distributions.
\par
Focusing on the accuracy of the predictive density, we rely on two distance-based metrics: the Continuous Rank Probability Score (CRPS) and the Mean Square Error (MSE). Let $y_p$ be the actual forecasting values with distribution $F_p$.
For every out-of-sample observation, the CRPS is defined as, $$\text{CRPS}(F_{l,p},y_{l,p})=\int_{-\infty}^{\infty}\(F\(\hat{y}_p\)-I_{\left\{\hat{y}_{l}\geq y_{l,p}\right\}}\)^2 d\hat{y}_{l},\quad l=1,\dots,L.$$ We compute the CRPS numerically, using the identity of \cite{Sze05}
$$\text{CRPS}\(F_{l,p},y_{l,p}\)=-\frac{1}{2}\mathbb E\left|\hat{Y}_{l}-\hat{Y}^{\prime}_{l,p}\right|-\mathbb E\left|\hat{Y}_l-y_{l,p}\right|,$$
were $\hat{Y}_l,\hat{Y}^{\prime}_l$ are independent replicates from the estimated (empirical) posterior predictive distribution. The MSE for the $l$-th out-of-sample observation is defined as 
\[\text{MSE}_{l}=\frac{1}{N}\sum^{N}_{i=1}\(y_{p,l}-\hat{y}_{l,i}\)^2,  \quad l=1,\dots,L,\] 
where $N$ is the MCMC sample size. We report the the CRPS and MSE for every prediction over all MCMC iterations and the average CRPS and MSE over all observations. The best model among its counterparts, is the one with the lowest CRPS and MSE.

\subsection{Overview of Methodology}
\begin{table}[!htb]
\setlength{\tabcolsep}{25pt}
\renewcommand{\arraystretch}{1.3}
\resizebox{\textwidth}{!}{%
\begin{tabular}{@{}LLLLL@{}}
\hline \clrone\multicolumn{5}{@{}C@{}}{} \\[-0.3cm]
\clrone
\b Model&\b Symbol & \b Predictors&\b Transition Probabilities&\b States\\
\hline\\[-0.4cm]
Non-Homogeneous Hidden Markov& (NHHM) &AR,X* & logistic/multinomial*&2-5\\
\clrone Homogeneous Hidden Markov&(HHM) &AR,X & constant&2-5\\
Markov Switching Random Walk&(MS-RW) &--- &constant &2\\
\clrone Kitchen Sink&(KS)&AR,X&--- &1\\
Linear Regression& (AR(5)) &AR& ---&1\\
\clrone Random Walk&(RW) &---&---&1\\
\hline 
\multicolumn{5}{@{}p{18cm}@{}}{*\small In the NHHM model, the predictors AR, X affect both the mean equation \emph{and} the transition probabilities. In all other models, they only affect the mean equation.}\vspace{-0.2cm}
\end{tabular}}
\caption{Summary of the models of this study. The first two columns show the model and its abbreviation, the third column shows the relevant predictors and the fourth column shows the assumed parametrization of the transition probabilities of each model. The last column shows the number of states that we consider.}
\label{tab:models}
\end{table}
Summing up, our methodology is the following. We study the performance of various Hidden Markov (HM) models with a fixed covariate set in explaining and predicting the cryptocurrency log-return series. In particular, we consider $4$ Non-Homogenous (NHHM) models with $m=2,\dots,5$ states, $4$ Homogeneous (HHM) models with $m=2,\dots, 5$ states and the 2-state Markov Switching Random Walk (MS-RW) model. \footnote{We omitted the $m$-state MS-RW models for $m>2$ since they do not offer an improved forecasting performance compared to the 2-state MS-RW model.}. To further benchmark our results, we also implement as single regime models the Random Walk (RW) model (i.e., a linear model with no covariates), the linear regression model with all the covariates and the autoregressive term, i.e., the Kitchen Sink (KS) model and an Autoregressive (AR(5)) model with lagged endogenous variables. This leads to 12 models for each coin that we summarize in \Cref{tab:models}. Then, we choose the model with the out-of-sample (predicting) performance based on the Continuous Rank Probability Score (CRPS) and Mean Squared Forecast Error (MSFE). Next, to study latent effects on the transition probabilities, we apply a reversible jump stochastic search algorithm on the multinomial regression predictors of the NHHM model with the best predicting performance. Finally, based on the included predictors and the in-sample performance of the best HM model, we give an economic interpretation of the hidden states for each cryptocurrency series. 


\section{Empirical Analysis}\label{sec:Empirical}
\subsection{Out-of-Sample analysis}
We asses the forecasting performance of the $m$-state, $m=2,\dots,5$, NHHM and HHM models, the 2-state MS-RW model and the benchmark models RW, KS and AR(5). The out-of-sample accuracy is assessed using a sequence of $L=30$ one-step-ahead predictive densities. In \Cref{tab:Forecasting Results}, we report the CRPS and MSE in parenthesis for 5 randomly chosen out-of-sample points (or horizons), $L=1,2,7,15$ and $30$. The last column reports the average scores over all the out-of-sample points. Through this exercise, parameter estimates are held fixed. 
\begin{table}[!htb]
\centering
\setlength{\tabcolsep}{15pt}
\resizebox{\textwidth}{!}{%
\begin{tabular}{@{}LRRRRRM}
\hline \clrone\multicolumn{7}{@{}C@{}}{} \\[-0.3cm]
\clrone
\multicolumn{7}{@{}C@{}}{\b Bitcoin}\\
\clrone
\b Horizon&\b 1&\b 2&\b 7&\b 15&\b 30& \multicolumn{1}{@{}r@{}}{\b Average}\\
\hline\\[-0.2cm]
NHHM$_2$&0.91 (8.79)&0.45 (8.20)& 0.65 (13.30) &0.82 (17.07)& 0.64 (14.80)&1.86 (28.80)\\
NHHM$_3$&1.00 (8.47)& 0.40 (8.82)&0.68 (14.13) & 0.88 (19.42)&0.61 (16.04)&1.85 (29.26)\\
NHHM$_4$&\b 0.71 (7.89)&\b 0.32 (6.37)&\b 0.58 (13.58) &\b 0.78 (15.62)&0.60 (15.54)&\b 1.78 (28.00)\\
NHHM$_5$&0.96 (15.33)&0.60 (14.27) &0.91 (25.92)&0.90 (18.97)& 0.62 (15.72)&1.85 (30.41)\\
\midrule
HHM$_2$&0.91 (9.67)& 0.52 (9.33)& 0.66 (14.61) & 0.86 (17.19) &0.67 (16.00)&1.87 (29.36)\\
HHM$_3$&0.93 (15.60)& 0.57 (15.67) & 0.78 (16.85) &0.83 (17.73)& 0.56 (15.72)&1.87 (30.02) \\
HHM$_4$&0.97 (14.61)& 0.62 (13.86)& 0.85 (23.06)&0.81 (16.73)&0.53 (14.85)&1.87 (29.43)\\
HHM$_5$&0.91 (14.26)& 0.49 (12.92) &0.75 (20.28) & 0.86 (15.13) &\b 0.50 (12.73)&1.83 (29.20)\\
\midrule
MS-RW&0.95 (9.80)&0.53 (8.93)&0.60 (14.00)&0.81 (16.05)&0.62 (15.50)&1.85 (29.50) \\
KS&1.10 (17.05)&0.95 (15.67)&0.90 (15.52)&1.09 (19.19)&0.92 (17.90)&1.92 (29.72)\\
AR(5)&1.12 (17.50)& 0.92 (15.35)&0.95 (15.69)&0.99 (16.20)&0.97 (18.12)&1.95 (30.09)\\
RW&1.20 (19.50)& 1.00 (15.85)&1.05 (14.89)&1.00 (16.15)&0.96 (18.85)&1.98 (31.12)\\[0.2cm]
\hline \clrone\multicolumn{7}{@{}C@{}}{} \\[-0.3cm]
\clrone
\multicolumn{7}{@{}C@{}}{\b Ether}\\ 
\hline\\[-0.2cm]
NHHM$_2$&1.60 (27.82)& 1.21 (27.83)&1.11 (29.22)& 0.89 (27.71)& 0.87 (27.09)&1.62 (31.51) \\
NHHM$_3$&1.59 (27.13)&1.15 (25.40)&1.15 (26.82)& 0.85 (25.80) & 0.93 (24.74)&1.60 (25.09)\\
NHHM$_4$&1.45 (20.45)&\b 1.04 (18.66)&\b 1.05 (19.21)&\b  0.69 (15.55)&\b 0.70 (15.95)&\b 1.56 (24.87)\\
NHHM$_5$&1.68 (31.57)&1.42 (30.12)&1.26 (31.59)&1.24 (28.08)&1.16 (27.43)&1.83 (37.12)\\
\midrule
HHM$_2$&1.60 (30.30)&1.29 (33.10)&1.15 (40.16)&1.10 (41.60)&1.07 (43.28)&1.75 (49.25)\\
HHM$_3$&1.67 (44.88)&1.48 (45.83)&1.24 (45.29)& 1.06 (41.83)&1.14 (41.41)&1.80 (51.47)\\
HHM$_4$&1.71 (39.94)& 1.35 (35.37) &1.22 (34.48) &1.21 (34.20) &1.14 (29.25)&1.81 (40.55)\\
HHM$_5$&1.97 (50.69) & 1.74 (46.94) & 1.57 (43.16)&1.58 (43.17)& 1.56 (44.20)&1.80 (43.42)\\
\midrule
MS-RW&\b 1.36 (28.40)& 1.20 (28.38)&1.11 (35.90)&1.07 (42.36)&1.22 (43.40)& 1.73 (48.86)\\
KS&1.97 (50.69) & 1.74 (46.94) & 1.57 (43.16)&1.58 (43.17)& 1.56 (44.20)&2.03 (52.00)\\
AR(5)&1.95 (50.45)&1.82 (47.89)&1.61 (42.79)& 1.49 (42.32) &1.62 (43.92)&2.04 (51.92)\\
RW&2.00 (51.65)&1.86 (48.03)&1.66 (43.87)& 1.48 (42.25) &1.66 (45.12)&2.15 (53.03)\\[0.2cm]
\hline \clrone\multicolumn{7}{@{}C@{}}{} \\[-0.3cm]
\clrone
\multicolumn{7}{@{}C@{}}{\b Ripple}\\
\hline\\[-0.2cm]
NHHM$_2$&0.98 (27.38)&1.24 (25.71)&0.81 (22.00)&0.87 (23.57)&0.75 (19.67)&1.54 (29.69)\\
NHHM$_3$&0.90 (26.45)&1.05 (23.91) &0.80 (23.86)&0.69 (22.34) &0.63 (16.92)&1.46 (28.67)\\
NHHM$_4$&\b 0.82 (24.50)& \b 0.99 (22.14)&\b 0.77 (21.70)&\b 0.60 (21.19)&\b 0.62 (18.69)& \b 1.39 (27.38)\\
NHHM$_5$&1.00 (31.50)&1.15 (27.87)&0.95 (25.34&0.71 (22.38) &0.73 (22.39)&1.60 (30.32)\\
\midrule
HHM$_2$&0.99 (28.00)&1.22 (24.40)&0.82 (21.88)&0.94 (22.80)&0.81 (23.37)&1.54 (29.16)\\
HHM$_3$&1.22 (37.94)&1.24 (33.45)&0.99 (30.69)&0.89 (29.15)&0.73 (28.56) &1.54 (36.40)\\
HHM$_4$&1.06 (36.42)&1.27 (26.97)&0.86 (27.55)&0.92 (26.35)&0.74 (24.18)&1.52 (32.96)\\
HHM$_5$&1.28 (33.41)&1.10 (29.07)&1.19 (38.27)&0.77 (27.34)&0.85 (27.16)&1.52 (33.68)\\
\midrule
MS-RW&1.25 (40.57)&1.09 (34.78)& 1.38 (41.25)&0.97 (39.33)&1.01 (39.05)&1.55 (44.93)\\
KS&2.08 (56.26) &2.00 (54.89) &1.64 (50.50)&1.85 (52.36)&1.66 (48.83)&2.09 (57.64)\\
AR(5)&1.96 (51.97)&2.05 (56.14) &1.74 (49.13)&1.72 (50.37)&1.67 (47.97)&2.10 (56.75)\\
RW&1.97 (52.02)&2.06 (56.08) &1.73 (48.45)&1.76 (52.01)&1.72 (50.11)&2.21 (58.73)\\
\bottomrule
\end{tabular}
}
\caption{Continuous Rank Probability Score and Mean Squared Error in parenthesis for all the competing models for the Bitcoin, Ether and Ripple series. The last column reports the average CRPS (MSE) over the whole sequence of 30 one-step ahead predictions. Bold values indicate the lowest CRPS values for each out-of-sample points.}
\label{tab:Forecasting Results}
\vspace{-0.3cm}
\end{table}

The first general observation is that all HM models significantly surpass the single regime RW, KS, AR(5) models, whereas all single regime models have similar forecasting performance. Notably, this is true even for the KS model which includes all the predictors and autoregressive terms. These results suggest that the HM models can identify time-varying parameterizations which lead to improved forecasting performance relatively to the single regime benchmarks. This is in line with previous findings on the necessity of incorporating the structural breaks and regime switches in modeling the BTC return series \citep{Thi18}. Moreover, it generalizes the evidence of time-varying effects in the BTC price series \citep{Me19} to the ETH and XRP price series.\par 
Concerning the predicting accuracy among the various HM models, we observe that the model with the best forecasting performance is the 4-state NHHM for all coins. This is based on the average CRPS and MSE scores (last column of \Cref{tab:Forecasting Results}). The 4-state NHHM model has the lowest CRPS for all the randomly chosen out-of-sample points. The only exceptions are the 30-th point in the BTC forecasting exercise, where it is outperformed by the 5-state HHM model, and the 1st point in the ETH forecasting exercise, where it is outperformed by the MS-RW model.\par
By collating the resulting MSEs for each point individually with the CRPS, we observe that in some forecasting horizons different models are found to outperform the best models derived by the CRPS. Even though this might seem contradictory, these differences are expected since the CRPS is more robust to outliers and more reliable when assessing the density forecasts (see \cite{Kok20} and references therein). Over all coins, the lowest CRPS and average MSE (best predicting accuracy) are achieved when predicting the BTC return series and the highest CRPS and average MSE (worst predictive accuracy) are achieved when predicting the ETH series.  
\par This forecasting exercise indicates that, besides the linear time-varying conditional correlations, there also exist non-linear correlations between the cryptocurrency return series and the considered predictor set. This is established by relaxing the hypothesis of constant transition probabilities which allows us to study the effects of the predictors on the series via a non-linear multinomial logistic relationship. The empirical results provide evidence that if we account for these more complex correlations between the observed cryptocurrency price series and the current predictor set, we obtain improved forecasting performance for the HM models under consideration.

\subsection{In-sample analysis of the models with the best predicting performance}
\begin{figure}[!htb]
\centering
\includegraphics[width=\linewidth,clip=true, trim = 4cm 1cm 3.5cm 1cm]{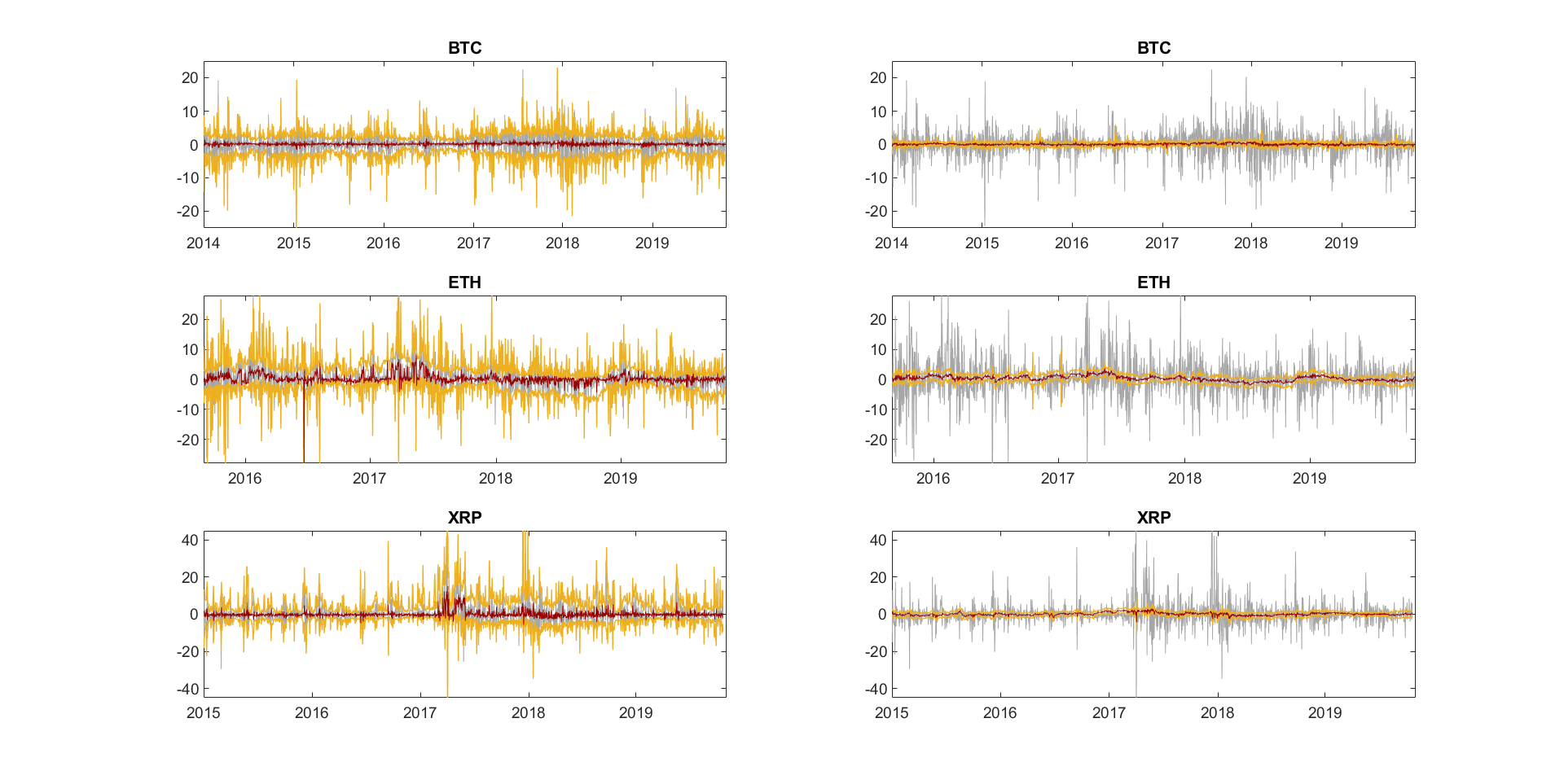}
\caption{Percentage return series (gray lines) of the posterior sample (replicated) empirical distributions for the Bitcoin series (first row), Ether series (second row) and Ripple series (third row). Yellow lines show the 0.5\% and 99.5\% quantiles of the estimated in-sample distributions and define the 1\% credibility region, whereas red lines show the estimated posterior median. Plots on the left are based on the estimated distributions via the 4-state Non-Homogeneous Hidden Markov (HM) model while plots on the right show the estimated distributions as derived from the Random Walk (RW) benchmark model.}
\label{fig:InSample}
\end{figure}

Based on the forecasting accuracy, we treat the 4-state NHHMs as our final model for further analysis for the BTC, ETH and XRP return series. Within our MCMC algorithm and at each iteration, we estimate an in-sample realization of the observed data. In particular, we use the in-sample estimations of the parameters and the states to reproduce the cryptocurrency percentage log return series, often referred to as \emph{replicated data} or \emph{within-sample predictions}, \cite{Ge03}. The derived realized distributions along with the observed series for each coin are shown in \Cref{fig:InSample}. 
The left column shows the in-sample replicated distributions of the aforementioned 4-state NHHM and the right column shows the replicated distributions of the RW benchmark for all coins. The gray lines show the observed percentage log-return series, the yellow lines show the fitted $0.5\%$-th and $99.5\%$-th quantiles of the estimated in-sample distribution and the red line shows the $50\%$-th quantile (median). By visual inspection, we observe that, by identifying the various volatility clusters, the 4-state NHHM model offers substantially improved in-sample performance compared to the in-sample performance of the RW model. The graphical proof of the good in-sample performance of the 4-state NHHM compared to the RW model is substantiated by \Cref{tab:Coverage} which shows the overall empirical coverage of the estimated quantile curves. The first and second rows report the proportion of the observed percentage log-returns that fall out of the empirical quantiles curves for the 4-state NHHM, and the in-sample MSE, respectively, for the 4-state NHHM, while the third and fourth rows report the proportion and in-sample MSE for the RW model.  

\begin{table}[!htb]
\setlength{\tabcolsep}{15pt}
\centering
\adjustbox{max width=\textwidth}{%
\begin{tabular}{@{}LLCCCC@{}}
\hline \clrone\multicolumn{5}{@{}C@{}}{} \\[-0.3cm]
\clrone
&&\b Bitcoin&\b Ether&\b Ripple\\
\hline\\[-0.3cm]
\b NHHM$_4$&Proportion&0.05 (121/2114) &0.06 (95/1506)&0.04 (79/1748)\\
&MSE&15.18&37.13&43.22\\
\midrule
\b RW &Proportion&0.75 (1601/2114)&0.74 (1114/1506)&0.72 (1255/1748)\\
&MSE&17.03&42.28&50.01\\
\bottomrule
\end{tabular}
}
\caption{Empirical coverage of the empirical in-sample distributions using the 4-state NHHM and the RW benchmark for the Bitcoin, Ether and Ripple percentage return series.}
\label{tab:Coverage}
\end{table}

\par \Cref{tab:Estimates} provides a gauge of what drives the documented predictability by showing the posterior mean estimates for the linear regression predictors for each state for BTC, ETH and XRP respectively. Predictors that fall into the 10\% credibility intervals are marked with asterisk. In addition to the linear regression estimates, the last column of \Cref{tab:Estimates} shows the posterior probabilities of inclusion for the predictors affecting the transition probabilities, as derived from the probabilistic variable selection scheme, i.e., the reversible jump stochastic search algorithm on the multinomial regression. Posterior probabilities of inclusion that exceed 0.4 are marked with bold fonts. \begin{table}[!htb]
\centering
\setlength{\tabcolsep}{27pt}
\begin{tabular}{@{}lS[table-format=2.2]S[table-format=1.2]S[table-format=1.2]S[table-format=1.2]S[table-format=1.2]@{}}
\hline\clrone\multicolumn{6}{@{}C@{}}{} \\[-0.3cm]
\clrone
\multicolumn{6}{@{}c@{}}{\b BTC}\\
\clrone
\multicolumn{1}{@{}L@{}}{\b Predictors}&\multicolumn{1}{@{}c@{}}{\b State 1}&\multicolumn{1}{@{}c@{}}{\b State 2}&\multicolumn{1}{@{}c@{}}{\b State 3}& \multicolumn{1}{@{}c@{}}{\b State 4}& \multicolumn{1}{@{}c@{}}{\b Probabilities}\\
\hline\\[-0.2cm]
\b Intercept&11.47&6.65&-0.80&0.02&0.00\\
\b EUR/USD&1.02&0.43&-0.15&-1.59&0.00\\
\b GBP/USD&0.06&0.17&-0.04$^{\ast}$&-1.26&0.00\\
\b CNY/USD&-0.14&-0.12&-0.15&-1.30&0.00\\
\b JPY/USD&-0.37$^{\ast}$&0.08&-0.16&-0.78&0.00\\
\b Gold&-0.09&-0.12&0.07&0.01&0.00\\
\b Oil&0.02&0.14&0.01&0.52&0.00\\
\b SP500&-6.73&-0.85&0.32&0.08&0.00\\
\b VIX&-4.37&-2.34&0.30&0.03&\b 1.00\\
\b TY&-2.07&-2.21&1.49&0.15&\b 0.90\\
\b Size&-0.01&-0.01&0.01&0.13&0.00\\
\b Hash&-0.01&-0.01&0.00&0.26&0.00\\
\b AR(1)&-0.04&0.00&0.03&0.94$^{\ast}$&0.00\\[0.2cm]
\hline \clrone\multicolumn{6}{@{}C@{}}{} \\[-0.3cm]
\clrone
\multicolumn{6}{@{}C@{}}{\b ETH}\\
\hline\\[-0.3cm]
\b Intercept&7.68&1.93&0.85&0.80&\b 1.00\\
\b EUR/USD&-0.14&0.10&0.64&0.81&\b 1.00\\
\b GBP/USD&1.78&0.51$^{\ast}$&2.35&2.14&0.00\\
\b CNY/USD&-0.40&-0.24&-0.18&0.31&0.00\\
\b JPY/USD&-0.14&0.31&1.49&1.54&0.00\\
\b Gold&-0.12&0.32$^{\ast}$&1.17&1.26&\b 0.82\\
\b Oil&-2.02$^{\ast}$&0.26$^{\ast}$&0.24&0.42&\b 1.00\\
\b SP500&-9.14&-3.52&1.62&1.91& 0.11\\
\b VIX&-2.54$^{\ast}$&-0.78$^{\ast}$&-1.01&-0.78&\b 1.00\\
\b TY&-13.45&3.75&1.45&0.31&\b 0.84\\
\b Size&0.07&0.21&0.10&0.03&0.00\\
\b Hash&0.03&0.00&0.17&0.12&0.00\\
\b AR(1)&-0.06&-0.17$^{\ast}$&0.74&0.82&0.00\\[0.2cm]
\hline \clrone\multicolumn{6}{@{}C@{}}{} \\[-0.3cm]
\clrone
\multicolumn{6}{@{}C@{}}{\b XRP}\\
\hline\\[-0.3cm]
\b Intercept&9.30&0.25&0.87&0.96&\b 1.00\\
\b EUR/USD&-1.88$^{\ast}$&-0.19&0.25$^{\ast}$&0.07&\b 1.00\\
\b GBP/USD&-4.03&-0.04&-0.25&-0.85&0.01\\
\b CNY/USD&3.31&0.26&0.13&0.48&0.02\\
\b JPY/USD&1.36&1.13&0.02&2.12&\b0.70\\
\b Gold&-1.44&-0.10&-0.01&-0.62&0.10\\
\b Oil&0.44&-0.07&-0.09&0.17$^{\ast}$&0.06\\
\b SP500&2.00&-7.06&-0.87&0.13&\b0.40\\
\b VIX&-1.88&-0.19&0.25&-0.85&\b 1.00\\
\b TY&-22.46&1.29&3.68&-0.08&\b 0.52\\
\b Size&0.07&0.21&0.10&0.03&0.00\\
\b AR(1)&0.44&-0.08&-0.10$^\ast$&0.18&0.00\\
\bottomrule
\end{tabular}
\caption{Posterior mean estimates of the 4-state Non-Homogeneous Hidden Markov model for the Bitcoin, Ether and Ripple percentage return series. The first column specifies the predictors. The second, third, fourth and fifth columns report the posterior mean estimates for each predictor at the first, second, third and fourth states respectively. The last column reports the posterior probabilities of inclusion for the predictors that affect the transition probabilities via the multinomial regression model. These probabilities are calculated by applying a stochastic search reversible jump algorithm within the MCMC scheme. Statistical significance at the 10\% level is denoted with $\ast$ and posterior probabilities exceeding 0.4 are marked with bold fonts.}
\label{tab:Estimates}
\vspace{-0.3cm}
\end{table}

\par We observe that the majority of the predictors are not statistically significant in the linear regression parametrization, especially for the BTC return series. At this point, it is important to stress that even when the coefficient of an explanatory variable is not statistically different from zero, this does not necessarily mean that the variable has no predictive power for the return series \citep{Pan15}. It is often the case that a variable that is insignificant in-sample has predictive out-of-sample power and vice versa. This argument is strengthened by our extensive experimental study --- results not reported here --- which shows that if we remove the insignificant predictors, the forecasting accuracy deteriorates. Furthermore, we observe that depending on the hidden state, the mean posterior estimates can be markedly different and even change their sign. Concerning the transition probabilities, we observe that there are several predictors with significant effects (high posterior probabilities of inclusion) that may vary between the three different coins. For instance, we observe that the commodity returns (Gold and Oil) affect only the transition probabilities of the ETH series while the SP500 returns affect the transition probabilities only for the XRP series. Among the predictors, the Volatility Index (VIX) and the Treasury Yield (TY) affect the transition probabilities for all series. In addition, we identify linear and non-linear connections with the exchange rates EUR/USD, GBP/USD, JPY/USD for all series but at different states. The existence of time- and state-conditional statistical significant predictors in all three cryptocurrencies conforms with and refines existing results concerning Bitcoin \citep{Kou19, Kur19}.

\subsection{Hidden States classification and interpretation}

\Cref{tab:Hidden states} reports the hidden state occupancies as the average time spent at each state $i$, $i=1,\dots,4$, the average returns and the corresponding standard deviation at each state.
\begin{table}[!htb]
\centering
\setlength{\tabcolsep}{15pt}
\begin{tabular}{@{}LLCS[table-format=4.2]S[table-format=1.2]@{}}
\hline\clrone\multicolumn{5}{@{}C@{}}{} \\[-0.3cm]
\clrone
\b Coin &\b State&\b Occupancies&\multicolumn{1}{@{}l@{}}{\b Average}&\multicolumn{1}{@{}r@{}}{\b Std}\\
\hline\\[-0.3cm]
\b BTC&1&0.35&-0.46&5.91\\
&2&0.30&0.49&2.49\\
&3&0.24&0.47&1.69\\
&4&0.11&0.02&0.53\\
\midrule
\b ETH&1&0.41&0.88&8.90\\
&2&0.54&-0.06&2.61\\
&3&0.02&0.16&1.13\\
&4&0.02&0.18&0.76\\
\midrule
\b XRP&1&0.17&3.59&14.98\\
&2&0.35&0.76&4.46\\
&3&0.45&0.41&1.77\\
&4&0.03&6.82&0.86\\
\bottomrule
\end{tabular}
\caption{Information on the states as derived form the experiments on the BTC, ETH, XRP return series. The first column reports the cryptocurrencies and the second column the different regimes (states). The third column reports the occupancies, i.e., the average time spend at each state. The fourth and fifth columns report the average returns and estimated standard deviation at each state.}
\label{tab:Hidden states}
\end{table}
 At a first glance, the hidden process identifies periods with different underlying volatilities for every coin, i.e., periods with high and low volatilities. In more detail, for the BTC series, it identifies periods with negative average returns and high volatilities (state 1), periods with positive returns and low volatility (states 2 and 3) and calm periods with average returns close to zero and very low volatility. This segmentation in the return series resembles the bear/turbulent (state 1) and bull (states 2 and 3) markets, while state 4 corresponds to a stable/calm regime. Furthermore, we observe that states 2 and 3 have similar (almost equal) average returns. The similar average returns and different volatility indicate that the hidden process segments the return series into two subseries with the same skewness but very different kurtosis \citep{Tim00}. 
\par
\begin{figure}[!htb]
\centering
\includegraphics[width=\linewidth, clip = true, trim = 4cm 1cm 3.5cm 1cm]{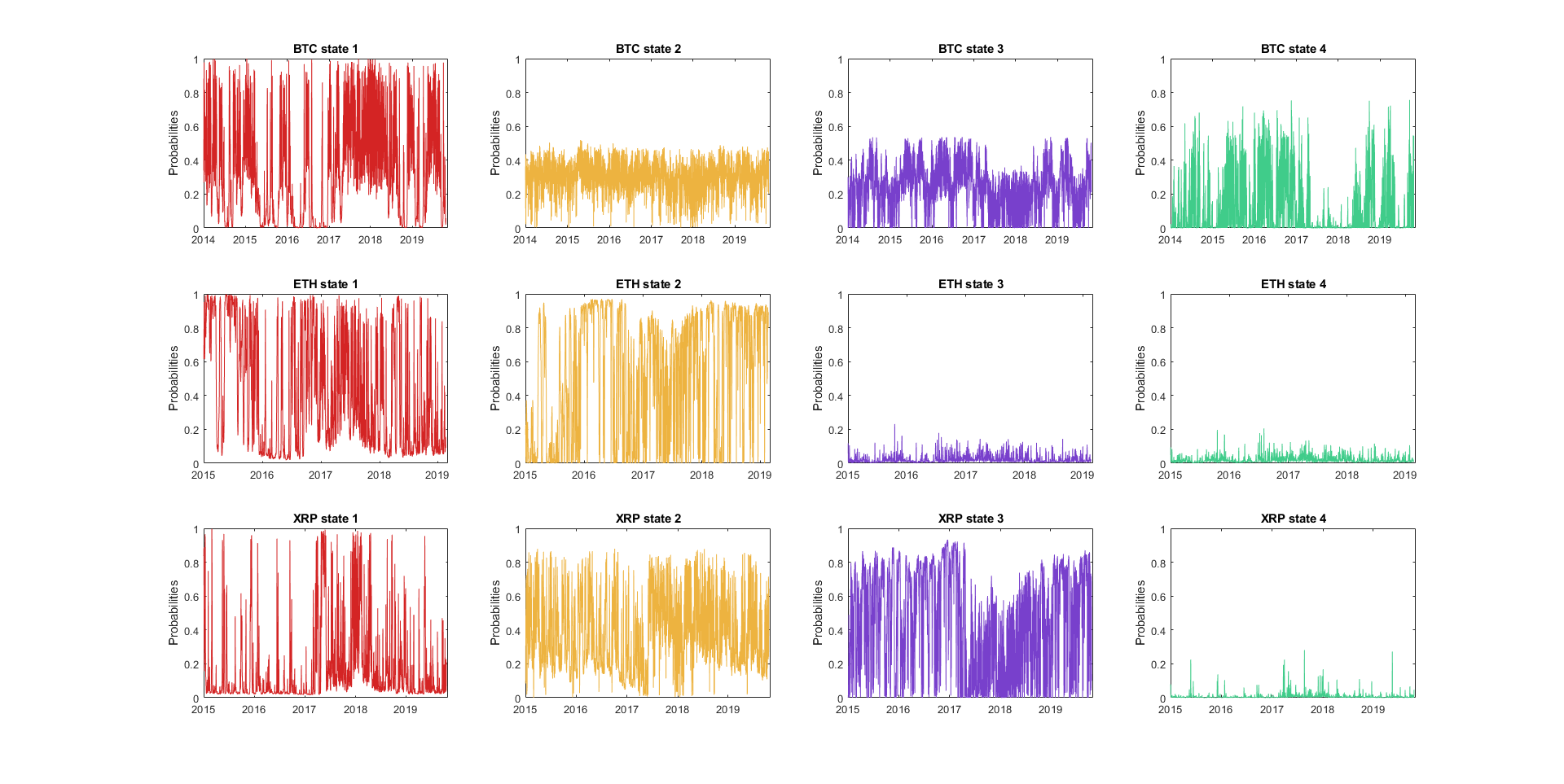}
\caption{Smoothed probabilities of being in state $m$, i.e., $P\(Z_t=m\mid y_1,\dots,y_T\)$, $m=1,\dots,4$ using the 4-state Non-Homogeneous Hidden Markov model. Columns 1 to 4 correspond to states 1 to 4 while rows 1 to 3 correspond to Bitcoin, Ether and Ripple, respectively.}
\label{fig:smoothed}
\end{figure}
\begin{figure}[!htb]
\centering
\includegraphics[width=\linewidth, clip = true, trim = 4cm 1cm 3.5cm 1cm]{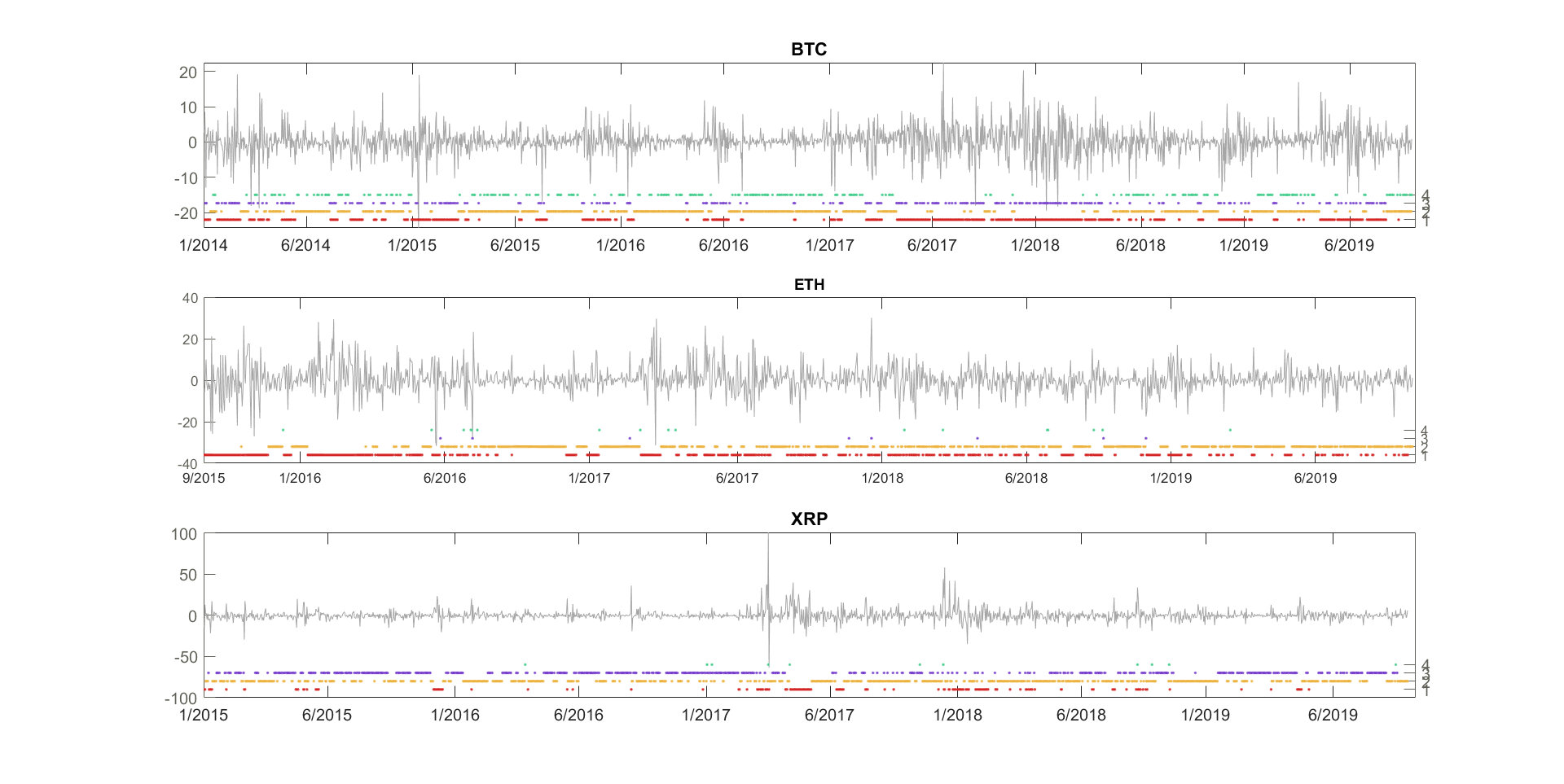}
\caption{Bitcoin (upper plot), Ether  (middle plot) and Ripple (lower plot) percentage return series conditionally on a realization of the hidden process. The hidden process is estimated using the 4-state NHHM on the  aforementioned cryptocurrencies' percentage return series. Red, yellow, purple and green dots indicate the time periods that the hidden process is at states 1, 2,3 and 4 respectively.}
\label{fig:returns}
\end{figure}

Concerning the ETH series, we observe that the highest mean returns occur in the state with the highest volatility. The hidden process alternates between a high volatility and a low volatility regime with almost zero average returns (states 1 and 2 respectively). This occurs for almost $95\%$ of the studied time frame. States 3 and 4 serve as auxiliary states with low occupancies and almost equal average returns but different volatilities. Finally, the hidden process in the XRP series spends most of the time ($80\%$) in the high volatility regime 2 and the low volatility regime 3. We also observe that state 1 has extremely high average returns compared to the returns of states 2 and 3 but is associated with very high risk (high volatility) as a trade-off. Lastly, hidden state 4 serves as an auxiliary state with low occupancy, capturing the extreme values (outliers) of the return series. 
 The information in \Cref{tab:Hidden states} can be visualized in \Cref{fig:smoothed,fig:returns} which depict the state-switching dynamics of the three cryptocurrencies according to their hidden state classification. 
\Cref{fig:smoothed} shows the estimated ex ante smoothed probabilities of each state for each time period, i.e., $P\(Z_t=m\mid y_1,\dots,y_T\)$, $m=1,\dots,4$ , over all coins. We observe that hidden state 1 for the BTC series, hidden states 1 and 2 for the ETH series and hidden states 1,2 and 3 for the Ripple series occur with high probability. The identification of these particular hidden states is sound with low probabilities of misclassification. However, we get mixed insights on the occurrence of the other states which are neither high nor low. \Cref{fig:returns} illustrates the Bitcoin (upper plot), Ether (middle plot) and Ripple (lower plot) returns conditionally on the a realization of the hidden process using the 4-state NHHM model. Gray lines correspond to the percentage log returns, while red, yellow, purple and green dots correspond to the times that the hidden state was in states 1,2,3 and 4, respectively. These graphical representations serve as an easy way to visualize the evolution of the hidden process in reference with the returns for each coin. While frequent alternations between the hidden states are prevalent in all three time-series, the transitional patterns are markedly different. For instance, in the BTC return series, there exist frequent alternations between states 1 and 3 and between 2 and 4, while in the ETH and XRP series they are between states 1 and 2 and between 2 and 3, respectively.

\section{Conclusion}\label{sec:Conclusion}
In this work, we modeled the return series of the three largest (in terms of market capitalization) cryptocurrencies, Bitcoin, Ether and Ripple using a Hidden Markov framework. We employed a multi-state Bayesian Hidden Markov methodology with a predefined set of financial and cryptocurrency specific predictors to capture the time-varying characteristics and heteroskedasticity of the cryptocurrencies' return series. The employed methodology is motivated by the existing evidence of structural breaks and regime/states switches in the cryptocurrency series. In line with the literature, we chose the best model among 9 different Hidden Markov models --- the standard Markov-Switching Random Walk (MS-RW) model, the Homogeneous Hidden Markov (HHM) and the Non-Homogeneous Hidden Markov (NHHM) models with up to five hidden states --- and 3 single regime models --- the Random Walk (RW) model, the linear AutoRegressive (AR) model and the Kitchen Sink (KS) model --- based on their out-of-sample predictive ability.
\par 
The out-of-sample forecasting exercise indicated that the 4-states NHHM model has the best forecasting performance for all three series with significant improvements over the single regime models. In addition, the regime switches reveal time-varying connections of traditional financial and economic predictors with cryptocurrency returns. In particular, we identified predictors affecting the series linearly, i.e., through the observed process, and non-linearly, i.e., through the unobserved process. From the complete set of twelve predictors, Bitcoin series is affected linearly and non-linearly by five predictors, while the Ether and Ripple series are affected by seven predictors. In general, only the US Treasury Yield and the CBOE stock market volatility index, VIX, have predictive power on all three series. Turning to the in-sample analysis, the 4-states NHHM model segments the return series into four subseries with distinct state-switching dynamics and economic interpretation. For Bitcoin, we find that the most frequently occurring state (state 1) corresponds to a bear regime (i.e., negative returns and high volatility), states 2 and 3 correspond to bull regimes (positive returns and low volatility) but with different kurtosis and state 4 corresponds to a calm regime (returns close to 0 and low volatility). Regarding the Ether return series, we observe frequent alternations between the high and low volatility states (states 1 and 2), while states 3 and 4 serve as auxiliary states with low occupancies. Finally, state 1 of the Ripple series corresponds to periods with extremely high average returns but, as a trade off, also with high risk. States 2 and 3 are the states with the highest occupancies, while state 4 serves as an auxiliary state. 
\par
Our empirical findings demonstrate that multi-state Non-Homogeneous Hidden Markov models offer improved forecasts on all considered cryptocurrency return series. In addition, they provide evidence to support the existence of predictors with state-dependent, time-varying predicting power on the cryptocurrency series. These insights are particularly useful to investors, portfolio-managers and policy-makers. Importantly, they refine our understanding of the dynamic relationship between traditional financial markets and cryptocurrency returns, which is characterized by regime switches and frequent alternations. Thus, when taken into account, these dynamics can be leveraged to optimize investors' decisions regarding portfolio allocation and risk-diversification between conventional and cryptocurrency assets. Moreover, along with the absence of predictors that systematically predict cryptocurrency returns in all possible states, the frequent alternations suggest that interested parties ought to be cautious when using forecasts to inform their decisions. Finally, the study of regimes with markedly different economic characteristics and transition dynamics is important for regulatory authorities and cryptocurrency entrepreneurs. Specifically, this information should be used to improve the implementation of policies that seek to mitigate risks associated with cryptocurrency exchange rates and promote their wider public adoption.

\section*{Conflict of interest}
The authors declare that they have no conflict of interest.

\bibliography{cryptopricebib}
\bibliographystyle{harvard}\biboptions{authoryear}

\end{document}